\def\year{2020}\relax
\documentclass[letterpaper]{article} % DO NOT CHANGE THIS
\usepackage{aaai20}  % DO NOT CHANGE THIS
\usepackage{times}  % DO NOT CHANGE THIS
\usepackage{helvet} % DO NOT CHANGE THIS
\usepackage{courier}  % DO NOT CHANGE THIS
\usepackage[hyphens]{url}  % DO NOT CHANGE THIS
\usepackage{graphicx} % DO NOT CHANGE THIS
\usepackage{graphicx}  % DO NOT CHANGE THIS
\usepackage{listings}  % Blagojce
\usepackage{dirtree} % Blagojce
\usepackage{mathptmx}
\usepackage{hyphenat}
\usepackage[T1]{fontenc}  % access \textquotedbl
\usepackage{textcomp}     % access \textquotesingle
\usepackage{tcolorbox}
\usepackage{xcolor}
\usepackage{balance}

\newcommand{\ie}{{i.e.}\xspace}
\newcommand{\eg}{{e.g.}\xspace}
\newcommand{\cf}{{cf.}\xspace}

\newcommand{\vs}{{vs.}\xspace}

\newcommand{\name}{WikiHist.html\xspace}
\newcommand{\WP}{Wikipedia\xspace}
\newcommand{\MW}{MediaWiki\xspace}

\newcommand{\zenodoDOI}{\url{https://doi.org/10.5281/zenodo.3605388}}

\newcommand*\samethanks[1][\value{footnote}]{\footnotemark[#1]}

\newcommand{\cpt}[1]{\textit{#1}}

\newcommand{\Secref}[1]{Sec.~\ref{#1}}

\newcommand{\Tabref}[1]{Table~\ref{#1}}
\newcommand{\Figref}[1]{Fig.~\ref{#1}}

\newcommand{\hide}[1]{}

\newcommand{\xhdr}[1]{\vspace{1.7mm}\noindent{{\bf #1.}}}

\newcommand{\reducedstrut}{\vrule width 0pt height .9\ht\strutbox depth .9\dp\strutbox\relax}
\newcommand{\colbox}[2]{%
  \begingroup
  \setlength{\fboxsep}{0pt}%  
  \colorbox{#1}{\reducedstrut#2\/}%
  \endgroup
}

\usepackage[show]{chato-notes}
\title{\name: English Wikipedia's Full Revision History in HTML Format}

\author{Blagoj Mitrevski\thanks{Authors contributed equally.}\\
EPFL\\
blagoj.mitrevski@epf\/l.ch
\And
Tiziano Piccardi\samethanks{}\\
EPFL\\
tiziano.piccardi@epf\/l.ch
\And
Robert West\\
EPFL\\
robert.west@epf\/l.ch
}

\begin{document}

\maketitle

\begin{abstract}

\WP is written in the wikitext markup language.
When serving content, the \MW software that powers \WP parses wikitext to HTML, thereby inserting additional content by expanding macros (templates and modules).
Hence, researchers who intend to analyze \WP as seen by its readers should work with HTML, rather than wikitext.
Since \WP's revision history is publicly available exclusively in wikitext format, researchers have had to produce HTML themselves, typically by using \WP's REST API for ad-hoc wikitext\hyp to\hyp HTML parsing.
This approach, however,
(1)~does not scale to very large amounts of data and
(2)~does not correctly expand macros in historical article revisions.
We solve these problems by developing a parallelized architecture for parsing massive amounts of wikitext using local instances of \MW, enhanced with the capacity of correct historical macro expansion.
By deploying our system, we produce and release \name, English \WP's full revision history in HTML format.
We highlight the advantages of \name over raw wikitext in an empirical analysis of \WP's hyperlinks, showing that over half of the wiki links present in HTML are missing from raw wikitext, and that the missing links are important for user navigation.
Data and code are publicly available at \zenodoDOI.

\end{abstract}

\section{Introduction}

Wikipedia constitutes a dataset of primary importance for researchers across numerous subfields of the computational and social sciences, such as social network analysis, artificial intelligence, linguistics, natural language processing, social psychology, education, anthropology, political science, human--computer interaction, and cognitive science.
Among other reasons, this is due to \WP's size, its rich encyclopedic content, its collaborative, self\hyp organizing community of volunteers, and its free availability.

Anyone can edit articles on \WP, and every edit results in a new, distinct revision being stored in the respective article's history.
All historical revisions remain accessible via the article's \textit{View history} tab.

\xhdr{Wikitext and HTML}
Wikipedia is implemented as an instance of MediaWiki,%
\footnote{\url{https://www.mediawiki.org}}
a content management system written in PHP, built around a backend database that stores all information.
The content of articles is written and stored in a markup language called \textit{wikitext} (also known as \textit{wiki markup} or \textit{wikicode}).%
\footnote{\url{https://en.wikipedia.org/wiki/Help:Wikitext}}
When an article is requested from \WP's servers by a client, such as a Web browser or the \WP mobile app, \MW translates the article's wikitext source into HTML code that can be displayed by the client.
The process of translating wikitext to HTML is referred to as \textit{parsing}.
An example is given below, in \Figref{fig:example}.

\begin{figure}[h!]
{\small
\begin{flushleft}
\textbf{Wikitext:}
\textsf{
\colbox{yellow!40}{\textquotesingle\textquotesingle\textquotesingle{}Niue\textquotesingle\textquotesingle\textquotesingle{}}
(\colbox{red!20}{$\{\!\{$lang-niu$|$Niu\=e$\}\!\}$}) is an
\colbox{blue!20}{[[island country]]}.
}
\end{flushleft}
}

% \noindent
% produces the HTML output

{\small
\begin{flushleft}
\textbf{HTML:}
\textsf{
\colbox{yellow!40}{<b>Niue</b>}
(\colbox{red!20}{<a href="/wiki/Niuean\_language"}
\colbox{red!20}{title="Niuean language">Niuean</a>:}
\colbox{red!20}{<i~lang="niu">Niu\=e</i>}) is an
\colbox{blue!20}{<a~href="/wiki/Island\_country"}
\colbox{blue!20}{title="Island country">island country</a>}.
}
\end{flushleft}
}
\vspace{-3mm}
\caption{
Example of wikitext parsed to HTML.
}
\label{fig:example}
\end{figure}

%%%%%%%%%%%%%%%%%%%%%%%%%%%%%%%%%%%%%%%%
% For example, parsing the wikitext

% {\small
% \begin{flushleft}
% \textsf{
% \colbox{yellow!40}{\textquotesingle\textquotesingle\textquotesingle{}Niue\textquotesingle\textquotesingle\textquotesingle{}}
% (\colbox{red!20}{$\{\!\{$lang-niu$|$Niu\=e$\}\!\}$}) is an
% \colbox{blue!20}{[[island country]]}.
% }
% \end{flushleft}
% }

% \noindent
% produces the HTML output

% {\small
% \begin{flushleft}
% \textsf{
% \colbox{yellow!40}{<b>Niue</b>}
% (\colbox{red!20}{<a href="/wiki/Niuean\_language"}
% \colbox{red!20}{title="Niuean language">Niuean</a>:}
% \colbox{red!20}{<i~lang="niu">Niu\=e</i>}) is an
% \colbox{blue!20}{<a~href="/wiki/Island\_country"}
% \colbox{blue!20}{title="Island country">island country</a>}.
% }
% \end{flushleft}
% }
%%%%%%%%%%%%%%%%%%%%%%%%%%%%%%%%%%%%%%%%

\noindent
Wikitext provides concise constructs for formatting text (\eg, as bold, \cf\ 
% \colbox{yellow!40}{yellow span}
yellow span
in the example of \Figref{fig:example}), inserting hyperlinks
% (\cf\ \colbox{blue!20}{blue }\colbox{blue!20}{span}),
(\cf\ blue span),
tables, lists, images, etc.

\xhdr{Templates and modules}
One of the most powerful features of wikitext is the ability to define and invoke so-called \textit{templates}.
Templates are macros that are defined once (as wikitext snippets in wiki pages of their own), and when an article that invokes a template is parsed to HTML, the template is expanded, which can result in complex portions of HTML being inserted in the output.
For instance, the template \textit{lang-niu,} which can be used to mark text in the Niuean language, is defined in the \WP page \cpt{Template:lang-niu}, and an example of its usage is marked by the
% \colbox{red!20}{red span}
red span
in the example of \Figref{fig:example}.
Among many other things, the infoboxes appearing on the top right of many articles are also produced by templates.
Another kind of wikitext macro is called \textit{module.} Modules are used in a way similar to templates, but are defined by code in the Lua programming language,
rather than wikitext.

\xhdr{Researchers' need for HTML}
The presence of templates and modules means that the HTML version of a \WP article typically contains more, oftentimes substantially more, information than the original wikitext source from which the HTML output was produced.
For certain kinds of study, this may be acceptable; \eg, when researchers of natural language processing use \WP to train language models, all they need is a large representative text corpus, no matter whether it corresponds to \WP as seen by readers.
On the contrary, researchers who study the very question how \WP is consumed by readers cannot rely on wikitext alone. Studying wikitext instead of HTML would be to study something that regular users never saw.

Unfortunately, the official \WP dumps provided by the Wikimedia Foundation contain wikitext only, which has profound implications for the research community:
researchers working with the official dumps study a representation of \WP that differs from what is seen by readers.
To study what is actually seen by readers, one must study the HTML that is served by \WP.
And to study what was seen by readers in the past, one must study the HTML corresponding to historical revisions.
Consequently, it is common among researchers of \WP 
%%%%%%%%%%%%%%%%%%%%%%%%%%%%%%%%%%%%%%%%%%
% For CRC: cite WWW'20 paper
%%%%%%%%%%%%%%%%%%%%%%%%%%%%%%%%%%%%%%%%%%
\cite{DBLP:conf/www/DimitrovSLS17,WtWRW,DBLP:conf/www/SingerL0ZWSL17} to produce the HTML versions of \WP articles by passing wikitext from the official dumps to the Wikipedia REST API,%
\footnote{\url{https://en.wikipedia.org/w/api.php}}
which offers an endpoint for wikitext\hyp to\hyp HTML parsing.

\xhdr{Challenges}
This practice faces two main challenges:
\begin{enumerate}
    \item Processing time: Parsing even a single snapshot of full English \WP from wikitext to HTML via the \WP API takes about 5 days at maximum speed. Parsing the full history of all revisions (which would, \eg, be required for studying the evolution of \WP) is beyond reach using this approach.
    \item Accuracy: MediaWiki (the basis of the \WP API) does not allow for generating the exact HTML of historical article revisions, as it always uses the latest versions of all templates and modules, rather than the versions that were in place in the past. If a template was modified  (which happens frequently) between the time of an article revision and the time the API is invoked, the resulting HTML will be different from what readers actually saw.
\end{enumerate}

Given these difficulties, it is not surprising that the research community has frequently requested an HTML version of Wikipedia's dumps from the Wikimedia Foundation.%
\footnote{See, \eg,
\url{https://phabricator.wikimedia.org/T182351}.}

\xhdr{Dataset release: \name}
With the \name dataset introduced in this paper, we address this longstanding need and surmount the two aforementioned hurdles by releasing the complete revision history of English Wikipedia in HTML format.
We tackle the challenge of scale (challenge 1 above) by devising a highly optimized, parallel data processing pipeline that leverages locally installed \MW instances, rather than the remote \WP API, to parse nearly 1~TB (bzip2\hyp compressed) of historical wikitext, yielding about 7~TB (gzip\hyp compressed) of HTML.

We also solve the issue of inconsistent templates and modules (challenge 2 above) by amending the default \MW implementation with custom code that uses templates and modules in the exact versions that were active at the time of the article revisions in which they were invoked. This way, we approximate what an article looked like at any given time more closely than what is possible even with the official \WP API.

In addition to the data, we release a set of tools for facilitating bulk\hyp downloading of the data and retrieving revisions for specific articles.

\xhdr{Download location}
Both data and code can be accessed via \zenodoDOI.
% \url{https://github.com/epfl-dlab/WikiHist.html}

\xhdr{Paper structure}
In the remainder of this paper,
we first describe the \name dataset (\Secref{sec:Dataset description}) and then sketch the system we implemented for producing the data (\Secref{sec:System architecture and configuration}).
Next, we provide strong empirical reasons for using \name instead of raw wikitext (\Secref{sec:Advantages of HTML over wikitext}), by showing that over 50\% of all links among \WP articles are not present in wikitext but appear only when wikitext is parsed to HTML, and that these HTML-only links play an important role for user navigation, with click frequencies that are on average as high as those of links that also appear in wikitext before parsing to HTML.

%%%%%%%%%%%%%%%%%%%%%%%%%%%%%%%%%%%%%%%%%%%%%%%%%%%%%%%%%%%%%%%%%%%%%%%%%%%%%%%%%%%%%%%%%

\section{Dataset description}
\label{sec:Dataset description}

The \name{} dataset comprises three parts:
the bulk of the data consists of English Wikipedia's full revision history parsed to HTML (\Secref{sec:HTML revision history}), which is complemented by two tables that can aid researchers in their analyses, namely
a table of the creation dates of all articles (\Secref{sec:Article creation dates}) and
a table that allows for resolving redirects for any point in time (\Secref{sec:Redirect history}).
All three parts were generated from English Wikipedia's revision history in wikitext format in the version of 1~March 2019.
For reproducibility, we archive a copy of the wikitext input%
\footnote{Downloaded from \url{https://dumps.wikimedia.org/enwiki/}.}
alongside the HTML output.

\subsection{HTML revision history}
\label{sec:HTML revision history}

The main part of the dataset comprises the HTML content of 580M revisions of 5.8M articles generated from the full English Wikipedia history spanning 18 years from 1~January 2001 to 1~March 2019.
Boilerplate content such as page headers, footers, and navigation sidebars are not included in the HTML.
The dataset is 7~TB in size (gzip\hyp compressed).
% \todo{Replace XXX above. Are sizes compressed or uncompressed? - Blagojce: compressed, furthermore I think this is the date of the oldest revision in the dump 2001-01-21T02:12:21Z}

% Every item in the collection represents a directory of the dataset (see below),

\xhdr{Directory structure}
The wikitext revision history that we parsed to HTML consists of 558 bzip2\hyp compressed XML files, with naming pattern
\url{enwiki-20190301-pages-meta-history$1.xml-p$2p$3.bz2},
where \url{$1} ranges from 1 to 27, and \url{p$2p$3} indicates that the file contains revisions for pages with ids between \url{$2} and \url{$3}.
Our dataset mirrors this structure and contains one directory per original XML file, with the same name.
Each directory contains a collection of gzip\hyp compressed JSON files, each containing 1,000 HTML article revisions.
Since each original XML file contains on average 1.1M article revisions, there are around 1,100 JSON files in each of the 558 directories.
% , named \url{1000.json.gz}, \url{2000.json.gz}, etc.

% giving rise to the following directory structure:
% \vspace{10pt}
% \dirtree{%
% .1 /.
% .2 enwiki-20190301-p-m-history1.xml/.
% .3 1000.json.gz.
% .3 2000.json.gz.
% .3 ....
% .3 1100000.json.gz.
% .2 ....
% .2 enwiki-20190301-p-m-history558.xml/.
% .3 1000.json.gz.
% .3 2000.json.gz.
% .3 ....
% .3 1100000.json.gz.
% }
% \vspace{10pt}

% \todo{Blagojce: I have a remark about the explanation of the XML file names, I wrote them in an abbreviated form in order to save space, but the full names of the 558 files can be seen here: https://pastebin.com/E6ZZA8Yi . The p10p2062 in the name of the first file means that this file starts with the page with id10 and contains all pages until page with id2062 (assuming that no page with id10-id2062 was deleted from Wikipedia).}

% \todo{What about page\_id\_count.txt, which is available in every folder? We need to describe that, too! Blagojce: I could just remove it from IA, it was used to generate the metadata for the look-up script.}

\xhdr{File format}
Each row in the gzipped JSON files represents one article revision.
Rows are sorted by page id, and revisions of the same page are sorted by revision id.
As in the original wikitext dump, each article revision is stored in full, not merely as a diff from the previous revision.
In order to make \name{} a standalone dataset, we include all revision information from the original wikitext dump, the only difference being that we replace the revision's wikitext content with its parsed HTML version (and that we store the data in JSON rather than XML).

The schema therefore mirrors that of the original wikitext XML dumps,
\footnote{\url{https://www.mediawiki.org/w/index.php?title=Help:Export&oldid=3495724#Export_format}}
but for completeness we also summarize it in Table~\ref{table1}a.

\begin{table}[t]
\centering
\caption{JSON schemas of dataset.
All fields in HTML revision history are copied from wikitext dump, except \url{html}, which replaces the original \url{text}.
}

\bigskip

\resizebox{\columnwidth}{!}{ % If your table exceeds the column or page width, use this command to reduce it slightly
\begin{tabular}{l|l}

\multicolumn{2}{c}{\textbf{(a) HTML revision history (\Secref{sec:HTML revision history})}\smallskip}\\
\textit{Field name}      & \textit{Description}\\ \hline
\url{id}              & id of this revision\\
\url{parentid}        & id of revision modified by this revision\\
\url{timestamp}       & time when revision was made\\
\url{cont_username}  & username of contributor\\
\url{cont_id}        & id of contributor\\
\url{cont_ip}        & IP address of contributor\\
\url{comment}         & comment made by contributor\\
\url{model}           & content model (usually \url{wikitext})\\
\url{format}          & content format (usually \url{text/x-wiki})\\
\url{sha1}            & SHA-1 hash\\
\url{title}           & page title\\
\url{ns}           & namespace (always 0)\\
\url{page_id}        & page id\\
\url{redirect_title} & if page is redirect, title of target page\\
\url{html}            & revision content in HTML format\\

\hline
\multicolumn{2}{c}{\vspace{2mm}}\\
\multicolumn{2}{c}{\textbf{(b) Page creation times  (\Secref{sec:Article creation dates})}\smallskip}\\
\textit{Field name}      & \textit{Description}\\ \hline
\url{page_id}        & page id\\
\url{title}           & page title\\
\url{ns}           & namespace (0 for articles)\\
\url{timestamp}       & time when page was created\\

\hline
\multicolumn{2}{c}{\vspace{2mm}}\\
\multicolumn{2}{c}{\textbf{(c) Redirect history (\Secref{sec:Redirect history})}\smallskip}\\
\textit{Field name}      & \textit{Description}\\ \hline
\url{page_id}        & page id of redirect source\\
\url{title}           & page title of redirect source\\
\url{ns}           & namespace (0 for articles)\\
\url{revision_id}        & revision id of redirect source\\
\url{timestamp}       & time at which redirect became active\\
\url{redirect}       & page title of redirect target (in 1st item\\
                     & of array; 2nd item can be ignored)\\
\hline

\end{tabular}
}
\label{table1}
\end{table}

\xhdr{Hyperlinks}
In live Wikipedia, hyperlinks between articles appear either as blue or as red.
Blue links point to articles that already exist (\eg, \url{/wiki/Niue}),
whereas red links indicate that the target article does not exist yet 
(\eg, \url{/w/index.php?title=Brdlbrmpft&action=edit&redlink=1}).
This distinction is not made in the wikitext source, where all links appear in identical format (\eg, \url{[[Niue]]}, \url{[[Brdlbrmpft]]}), but only when the respective article is requested by a client and parsed to HTML.
As the existence of articles changes with time, we decided to not distinguish between blue and red links in the raw data and render all links as red by default.
In order to enable researchers to determine, for a specific point in time, whether a link appeared as blue or red and what the hyperlink network looked like at that time, we also provide the two complementary datasets described next.

% In the dataset, all hyperlinks to English Wikipedia articles appear as \texttt{index.php?title=XYZ}.
% We do not distinguish between blue links (pointing to existing target articles) and red links (pointing to target articles that do not exist yet).

% Note that this is different from live Wikipedia, where the format is \texttt{/wiki/XYZ} for 
% Given the architecture of the system that we used to generate the dataset, the format of the links does not include any rewrite rule, and they are expressed in the raw form: \textit{index.php?title=[name]}

\subsection{Page creation times}
\label{sec:Article creation dates}

% Together with the primary dataset, we release some complementary datasets that can help with specific tasks.

% On the live version of Wikipedia, it is possible to infer from every link of the HTML version if the destination article exists or not. During the conversion from Wikitext to HTML, MediaWiki checks the presence of the article in the database, and if it is missing, it marks the tag with a CSS class. With the default skin, these links are the ones that appear \textit{red} in the text.

% Since the presence of articles changes in time, the released dataset does not make this distinction, and the links do not have a property to recognize if they point to missing articles.

% To enable studies focused on the evolution of the link network, we distribute a complementary dataset\footnote{This dataset is available as ``page\_creation\_timestamp.json.gz''} that summarises the creation time of each of the articles available in Wikipedia, redirects included.

The lookup file \url{page_creation_times.json.gz} (schema in \Tabref{table1}b) specifies the creation time of each English Wikipedia page.
% , including redirects.
To determine if a link to a target article $A$ was blue or red at time $t$ (\cf\ \Secref{sec:HTML revision history}), it suffices to look up $A$ in this file.
If $A$ was created after time $t$ or if it does not appear in the file, the link was red at time $t$; otherwise it was blue.

% {
%   "page_id": "5135781",
%   "title": "Gigi (song)",
%   "ns": "0",
%   "timestamp": "2006-05-14T01:42:05Z"
% }

% The state of the page links in a specific instant in time can be obtained with the rules:
% \begin{itemize}
%     \item If the page is missing, the link is \textit{red} and article was not present in March 2019
%     \item If the article is present, the state can be inferred by comparing the creation time of the articles with the date of interest.
% \end{itemize}

% \begin{table*}[t]
% \centering
% \caption{The link structure in the HTML.}\smallskip
% %\resizebox{0.95\textwidth}{!}{ % If your table exceeds the column or page width, use this command to reduce it slightly
% \begin{tabular}{l|l|l}
% Type     & Format                     & Description                                   \\ \hline
% Internal & /index.php?title={[}name{]} & Internal link to a page with title {[}name{]} \\
% Interal  & /index.php/File:{[}name{]} & Internal link to a file with name {[}name{]} \\
% External & http://.*                  & External to page using HTTP protocol          \\
% External & https://.*                 & External to page using HTTPS protocol        
% \end{tabular}
% %}
% \label{table2}
% \mynote{We don't need this table}
% \end{table*}

\subsection{Redirect history}
\label{sec:Redirect history}

Wikipedia contains numerous redirects, i.e., pages without any content of their own whose sole purpose is to forward traffic to a synonymous page.
For instance, \cpt{Niue Island} redirects to \cpt{Niue}.
Link occurrences in the wikitext dumps, as well as our derived HTML dumps, do not specify whether they point to a proper article or to a redirect.
Rather, redirects need to be explicitly resolved by researchers themselves, a step that is complicated by the fact that redirect targets may change over time.
Since redirect resolution is crucial for analyzing Wikipedia's hyperlink network, we facilitate this step by also releasing the full redirect history as a supplementary dataset: the file \url{redirect_history.json.gz} (schema in \Tabref{table1}c) specifies all revisions corresponding to redirects, as well as the target page to which the respective page redirected at the time of the revision.

% {
%   "title": "G-d",
%   "ns": "0",
%   "page_id": "12698",
%   "revision_id": "68144862",
%   "timestamp": "2006-08-07T06:27:09Z",
%   "redirect": [
%     "Names of God in Judaism#Laws_of_writing_divine_names",
%     ""
%   ]
% }

% In Wikipedia, editors can create articles with redirect behavior. Although these articles belong to the same namespace of the regular pages, they are used to keep the navigation consistent in situations where the article is renamed. In practice, they are articles with a special tag pointing to the destination page, and they have an essential role for analysis when the links between pages play an important role.

% A critical aspect of redirects it that they can evolve in time. For example, the redirect \textit{Gipsy} had for some time the destination article \textit{Roma and Sinti} before changing to \textit{Romani people} when this article was created.

% This dataset, distributed in Newline delimited JSON format, enables the analysis of links network, giving a handy way to resolve the actual destination page at any instant in time.
% In particular, this dataset can be used with the following rules:

% \begin{itemize}
%     \item If the row does not exit, the article is not a redirect
%     \item If one of multiple rows exist, the creation time can be used to determine what the destination was at the date of interest
% \end{itemize}

\subsection{Limitation: deleted pages, templates, modules}
\label{sec:Limitation}

Wikipedia's wikitext dump contains all historical revisions of all pages that still existed at the time the dump was created.
It does not, however, contain any information on pages that were deleted before the dump was created.
In other words, when a page is deleted, its entire history is purged.
Therefore, since \name is derived from a wikitext dump, deleted pages are not included in \name either.

When using \name to reconstruct a past state of \WP, this can lead to subtle inaccuracies.
For instance, it follows that the rule of \Secref{sec:Article creation dates} for deciding whether a link was blue or red at time $t$ will incorrectly tag a link $(u,v)$ as red if $v$ existed at time $t$ but was deleted before 1~March 2019 (the date of the wikitext dump that we used).
Although such inconsistencies are exceedingly rare in practice, researchers using \name should be aware of them.

Since \MW handles templates and Lua modules (together referred to as \textit{macros} in the remainder of this section) the same way it treats articles (they are normal wiki pages, marked only by a prefix \textit{Template:} or \textit{Module:}), deleted macros are not available in the revision history either.
It follows that a deleted macro cannot be processed, even when parsing a revision created at a time before the macro was deleted.
This leads to unparsed wikitext remaining in the HTML output in the case of templates,
and to error messages being inserted into the HTML output in the case of Lua modules.

In some cases, we observed that editors deleted a macro and created it again with the same name later. This action introduces the problem of losing the revision history of the macro before its second creation. In such cases, we assume that the oldest macro revision available approximates best how the macro looked before its deletion and use that version when parsing article revisions written before the macro was deleted.

We emphasize that the limitation of deleted pages, templates, and modules is not introduced by our parsing process.
Rather, it is inherited from Wikipedia's deliberate policy of permanently deleting the entire history of deleted pages.
Neither can the limitation be avoided by using the \WP API to parse old wikitext revisions; the same inconsistencies and error messages would ensue.
On the contrary, \name produces strictly more accurate approximations of the HTML appearance of historical revisions than the \WP API, for the API always uses the latest revision of all templates and modules, rather than the revision that was actually in use at the time of the article revision by which it was invoked.

%%%%%%%%%%%%%%%%%%%%%%%%%%%%%%%%%%%%%%%%%%%%%%%%%%%%%%%%%%%%%%%%%%%%%%%%%%%%%%%%%%%%%%%%%

\section{System architecture and configuration}
\label{sec:System architecture and configuration}

\WP runs on \MW, a content management system built around a backend database that stores all information on pages, revisions, users, templates, modules, etc.
In this project we only require one core functionality: parsing article content from wikitext to HTML.
In \MW's intended use case, parsing is performed on demand, whenever a page is requested by a Web client.
Our use case, on the contrary, consists in bulk\hyp parsing a very large number of revisions.
Since \MW was not built for such bulk\hyp parsing, the massive scale of our problem requires a carefully designed system architecture.

\begin{figure}[t]
\centering
\includegraphics[width=0.9\columnwidth]{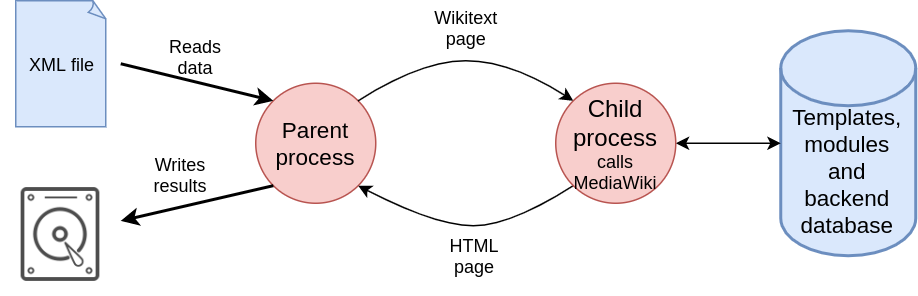} % Reduce the figure size so that it is slightly narrower than the column. Don't use precise values for figure width.This setup will avoid overfull boxes. 
\caption{Architecture for parsing \WP's revision history from wikitext to HTML.}
\label{fig2}
\end{figure}

\xhdr{System overview}
Our solution is schematically summarized in \Figref{fig2}.
As mentioned in \Secref{sec:HTML revision history}, the input to the parsing process consists of the hundreds of XML files that make up English \WP's full revision history in wikitext format.
Our system processes the XML files in parallel, each in a separate parent process running on a CPU core of its own.
Parent processes read the data from disk (in a streaming fashion using a SAX XML parser) and spawn child processes that parse the article contents from wikitext to HTML.
% \todo{Is the above statement correct? Blagojce: There could be more parent processes on a single machine. Actually there are as many parent processes as available CPU cores meaning that on 48 cores CPU, concurrently we were processing 48 XML files, for each file one parent and one child process at a time. }
Each child process has access to its own dedicated \MW instance.
The parent processes collect the HTML results from the child processes and write them back to disk.
Although this architecture is straightforward in principle, several subtleties need to be handled, described next.

% \begin{figure}[t]
% \centering
% \includegraphics[width=0.9\columnwidth]{fetch_templates.png} % Reduce the figure size so that it is slightly narrower than the column. Don't use precise values for figure width.This setup will avoid overfull boxes. 
% \caption{The logic of fetching templates, example.}
% \label{fig1}
% \end{figure}

\xhdr{Template and module expansion}
Wikitext frequently invokes macros (templates and modules) that need to be expanded when parsing to HTML.
Since macros may (and frequently do) themselves change over time, it is important to use the version that was active at the time of the article revision that is being parsed, given that we aim to reconstruct the HTML as it appeared at the time of the article revision.
\MW unfortunately does not provide such a retroactive macro expansion mechanism, but instead always uses the latest available version of each macro.
We therefore provide a workaround ourselves, by implementing an interceptor that, every time a macro is expanded, selects the historically correct macro version based on the revision date of the page being parsed, and returns that macro version to the parser instead of the default, most recent version.%
\footnote{
To support this procedure, the caching mechanisms of MediaWiki must be turned off, which introduces significant latency.
}
More precisely, we select the most recent macro version that is older than the article revision being parsed.

\xhdr{MediaWiki version}
Not only templates and modules, but also the \MW software itself has changed over time, so in principle the same wikitext might have resulted in different HTML outputs at different times.
To strictly reproduce the exact HTML served by \WP at a given time, one would need to use the \MW version deployed by \WP at that time.
Juggling multiple versions of \MW would, however, severely complicate matters, so we started by consulting the Internet Archive Wayback Machine%
\footnote{\url{https://archive.org/web/}}
in order to compare identical article revisions in different HTML snapshots taken at times between which live \WP's \MW version changed.
Screening numerous revisions this way, we found no noticeable differences in the HTML produced by different \MW versions and therefore conclude that it is safe to use one single \MW version for all revisions.
In particular, we use the latest long-term support version of \MW, 1.31.%
\footnote{\url{https://www.mediawiki.org/wiki/MediaWiki_1.31}}

\xhdr{Parser extensions}
\MW offers numerous extensions, but not all extensions used by live \WP are pre-installed in \MW's default configuration.
We therefore manually installed all those extensions (including their dependencies) that are necessary to reproduce live Wikipedia's parsing behavior.
In particular, we mention two crucial parser extensions:
\textit{ParserFunctions},\footnote{\url{https://www.mediawiki.org/wiki/Extension:ParserFunctions}}
which allows for conditional clauses in wikitext, and
\textit{Scribunto},\footnote{\url{https://www.mediawiki.org/wiki/Extension:Scribunto}}
the extension that enables the usage of Lua modules in wikitext.
% (modules are saved as regular \WP pages and can be invoked by articles in a way similar to templates).

% \begin{itemize}
%     \item ParserFunctions: This extension enriches the parser with logical and string functionalities: i.e. it makes possible to write conditional clauses in the Wikitext.
%     \item Scribunto: This extension allows the embedding Lua scripts in the Wikitext. Lua is a scripting language designed for embedded use in application. With the functionalities of the Scribunto extensions, contributors are enabled to write Lua scripts, save them as pages, and then invoke these scripts in articles similarly as invoking templates. This extension allows for more complex transclusion of pages in the articles and using scripts for generating parts of the articles.
% \end{itemize}

% The downloadable version of MediaWiki offers some pre-installed extensions, but not the full set currently deployed on Wikipedia.
% After downloading and setting up MediaWiki, we install all the parser hooks extensions and resolve all their dependencies to reproduce the working environment of Wikipedia.

\xhdr{Database connectivity}
By design, \MW instances cannot run without a persistent database connection.
However, given that
(1)~wikitext\hyp to\hyp HTML parsing is the only functionality we require,
(2)~the input to be parsed comes directly from a wikitext dump rather than the database, and
(3)~we intercept template and module lookups with custom code (see above),
we never actually need to touch the \MW database.
Hence we need not populate the database with any data (but we still need to create empty dummy tables in order to prevent \MW from throwing errors).

\xhdr{Scaling up}
Given the amount of wikitext in the full revision history, parallelization is key when parsing it.
We explored multiple common solutions for scaling up, including Spark and Yarn, but none of them satisfied all our requirements.
Therefore, we instead settled on a custom, highly\hyp optimized implementation based on Docker%
\footnote{\url{https://en.wikipedia.org/w/index.php?title=Docker_(software)&oldid=934492701}}
containers:
we bundle the modified \MW installation alongside the required MySQL database into a standalone Docker container and ship it to each machine involved in the data processing.

\xhdr{Failure handling}
Failures can happen during the parsing process for multiple reasons, including malformed wikitext, memory issues, etc.
Detecting such failures is not easy in \MW's PHP implementation: in case of an error it calls the \texttt{die} function, which in turn interrupts the process without raising an exception.
As a workaround, the parent processes (one per XML file; see above) are also responsible for monitoring the status of the child processes: whenever one of them fails, the event is detected and logged.
By using these logs, processing of the failure\hyp causing revisions can be resumed later, after writing custom code for recognizing problematic wikitext and programmatically fixing it before sending it to the parser.
Our deployed and released code incorporates all such fixes made during development runs.

\xhdr{Computation cost}
We used 4 high-end servers with 48 cores and 256~GB of RAM each. Each core ran one parent and one child process at a time.
In this setup, parsing English \WP's full revision history from wikitext to HTML took 42 days and, at a price of CHF~8.70 per server per day,
cost a total of CHF~1,462.

\section{Advantages of HTML over wikitext}
\label{sec:Advantages of HTML over wikitext}

Our motivation for taking on the considerable effort of parsing \WP's entire revision history from wikitext to HTML was that raw wikitext can only provide an approximation of the full information available in a Wikipedia article, primarily because the process of parsing wikitext to HTML tends to pull in information implicit in external templates and modules that are invoked by the wikitext.

In this section, we illustrate the shortcomings of wikitext by showing that a large fraction of the hyperlinks apparent in the parsed HTML versions of \WP articles are not visible in wikitext, thus providing researchers with a strong argument for using \name instead of raw wikitext dumps whenever their analyses require them to account for all hyperlinks seen by readers
\cite{dimitrov2016visual,DBLP:conf/www/DimitrovSLS17,paranjape2016improving,west2012human}.

\xhdr{Prevalence of HTML-only links over time}
First we quantify the difference in the number of links that can be extracted from the wikitext \vs HTML versions of the same article revisions.
To be able to determine whether the difference has increased or decreased with time, we study the 10 years between 2010 and 2019.
In order to eliminate article age as a potential confound, we focus on the 404K articles created in 2009.
% To quantify the difference in the number of links that can be extracted from wikitext \vs HTML, we start with the 404K English articles created in 2009 and
For each article created in 2009, we study 10 revisions, viz.\ the revisions available at the start of each year between 2010 and 2019.
For each revision, we extract and count internal links (pointing to other English Wikipedia articles) as well as external links (pointing elsewhere) in two ways: (1)~based on the raw wikitext, (2)~based on the HTML available in \name.%
\footnote{
As internal links, we consider only links pointing to articles in the main namespace and without prefixes, thus excluding talk pages, categories, etc.
We exclude self-loops.
In all analyses, if the same source links to the same target multiple times, we count the corresponding link only once.
% \todo{@Tiziano: just double-checking: you also excluded such links from HTML, correct? - YES}
To extract internal links from wikitext, we used a regular expression crafted by \citeauthor{Consonni_Laniado_Montresor_2019} \shortcite{Consonni_Laniado_Montresor_2019}.
}

% select the revision of the articles available on January 1st of each year considered.
% To remove the age of the article as potential confounding, we select only the articles created in 2009.
% This filter gives us 10 samples for each of the 404k articles.
% Then, we extract the links by dividing them into internal and external, based on their destination.

% In the first case, \textbf{internal links}, we consider links that point to another Wikipedia article using the following rules:

% \begin{itemize}
    % \item HTML: we select links to Wikipedia respecting the pattern \textit{index.php?title=[name]}. We exclude self-loops and links to pages that are not in name-space 0.
    % \item WikiText: we use a regular expression \cite{Consonni_Laniado_Montresor_2019} to select links, and we limit to the ones without prefix to exclude Categories and Talk pages.
% \end{itemize}

% In the case of \textbf{external links} we use the following rules:

% \begin{itemize}
%     \item HTML: we select any link that points to a destination different from Wikipedia.
%     \item WikiText: we search with a regular expression of any sub-string that matches a potential URL.
% \end{itemize}

\begin{figure}[t]
\centering
\includegraphics[width=0.9\columnwidth]{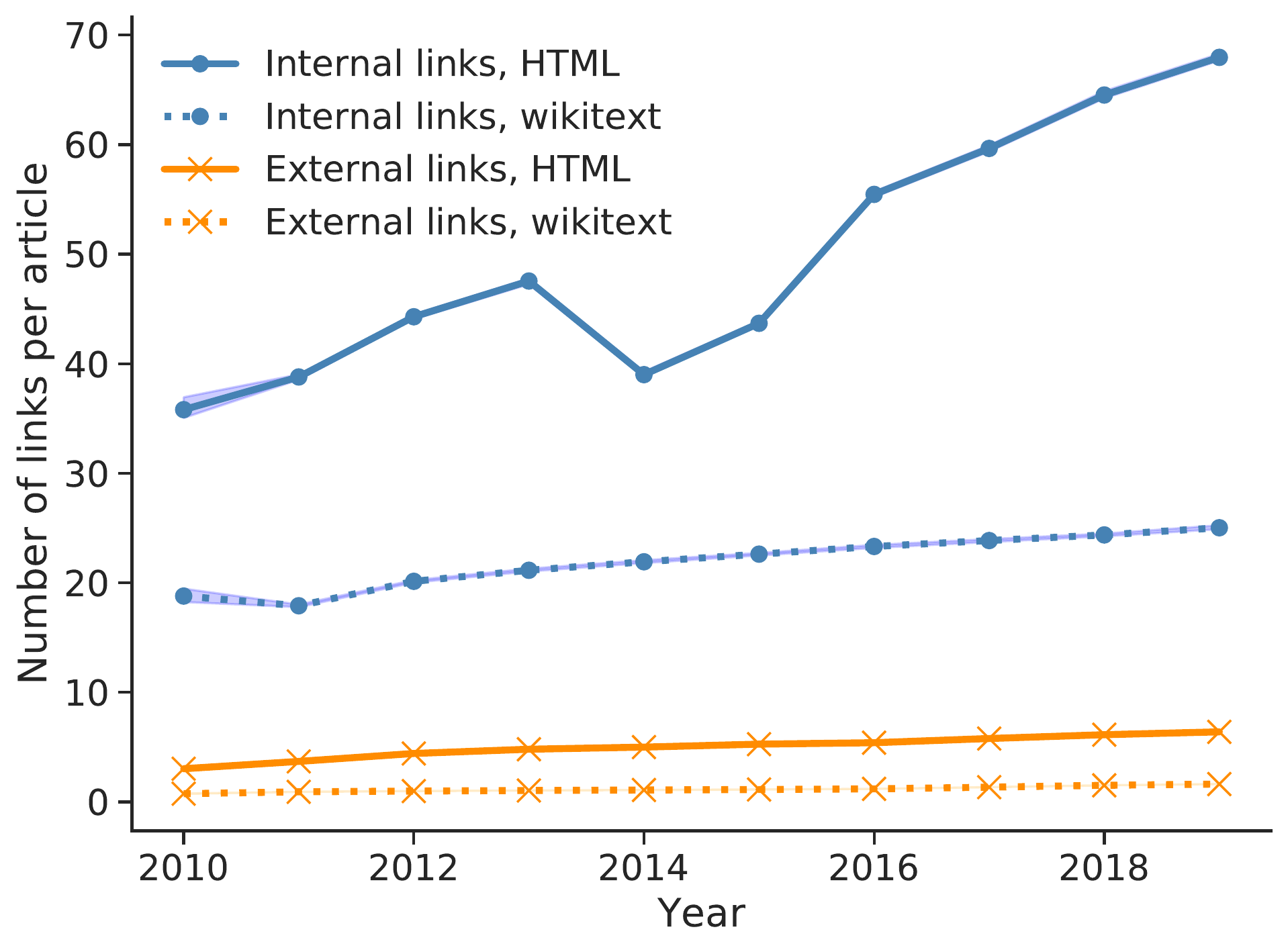} 
\caption{
Number of links extracted from wikitext and HTML, averaged over 404K articles created in 2009; 95\% error bands estimated via bootstrap resampling.
}
\label{average_links}
\end{figure}

\Figref{average_links} shows the number of links per year averaged over the 404K articles, revealing a large gap between wikitext and HTML.
The gap is significant (with non\hyp overlapping error bands) for both internal and external links, but is much wider for internal links.
Notably, for most years we can extract more than twice as many links from HTML as from raw wikitext, implying that researchers working with raw wikitext (presumably the majority of researchers at present) see less than half of all \WP{}\hyp internal links.

Via manual inspection we found that most of the links available in HTML only (henceforth ``HTML-only'' links) are generated by templates and Lua modules to enhance the navigation, \eg, in infoboxes on the top right of pages or as large collections of related links at the bottom of pages.%
\footnote{
The noticeable dip in 2014\slash 2015 of the number of internal links extracted from HTML (top, blue curve in \Figref{average_links}) was caused by the introduction of a then\hyp popular Lua module called \textit{HtmlBuilder}, which, among other things, automated the insertion of certain links during wikitext\hyp to\hyp HTML parsing.
The module was later deleted and could not be recovered (\cf\ \Secref{sec:Limitation}), thus leading to those links being unavailable in \name and therefore to an underestimation of the true number of links present during the time that \textit{HtmlBuilder} was active.
}
% \citeauthor{Consonni_Laniado_Montresor_2019} \shortcite{Consonni_Laniado_Montresor_2019} argue that links introduced by templates and modules are qualitatively different from links in the main article text

\xhdr{Popularity of HTML-only links}
Next we aim to determine how important HTML-only links are from a navigational perspective, operationalizing the importance of a link in terms of the frequency with which it is clicked by users of \WP.
If, for argument's sake, HTML-only links were never clicked by users, these links would be of little practical importance, and the necessity of working with \name rather than raw wikitext dumps would be less pronounced.
If, on the contrary, HTML-only links were clicked as frequently as links also available in wikitext, then researchers would see a particularly skewed picture by not observing over half of the available links.

Click frequency information is publicly available via the \WP Clickstream dataset,%
\footnote{\url{https://dumps.wikimedia.org/other/clickstream/}}
which counts, for all pairs of articles, the number of times users reached one article from the other via a click, excluding pairs with 10 or fewer clicks.
% \todo{@Tiziano: can you please verify the "10 or fewer" statement in the data? CORRECT}
We work with the January 2019 Clickstream release.%
\footnote{Since redirects have been resolved in the Clickstream, we also do so for links extracted from wikitext and HTML in this analysis.}

\begin{figure}[t]
\centering
\includegraphics[width=0.6\columnwidth]{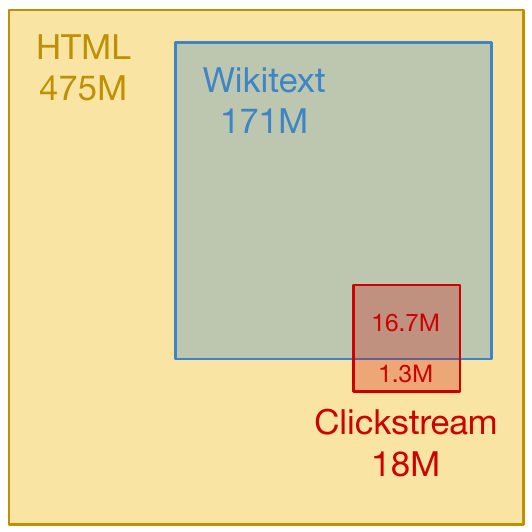} 
\caption{
Venn diagram of number of links in wikitext and HTML revisions of 1~January 2019, and in Clickstream release of January 2019.
% (\ie, over 10 clicks in January 2019).
}
\label{html_venn}
\end{figure}

The situation is summarized as a Venn diagram in \Figref{html_venn}.
On 1~January 2019, there were 475M internal links in \name (extracted from 5.8M articles).
Out of these, only 171M (36\%) are also present in wikitext, and 18M (3.8\%) are present in the Clickstream (\ie, were clicked over 10 times in January 2019).
Strikingly, out of the 18M links present in the Clickstream, 1.3M (7.2\%) cannot be found in wikitext, accounting for 6.1\% of all article\hyp to\hyp article clicks recorded in the Clickstream.
That is, joining Clickstream statistics with the contents of the respective articles is not fully feasible when working with raw wikitext. With \name, it is. 

% To further understand the importance of the 1.3M Clickstream links missing from the wikitext, we next ask how popular the missing links are, relative to the set of all links available in HTML.

We now move to quantifying the navigational importance of the 1.3M Clickstream links available in HTML only, relative to the set of all 18M Clickstream links available in HTML.
(In this analysis, we consider only the 18M links present in the Clickstream.)
For each of the 405K articles containing at least one HTML-only link, we sort all links extracted from \name by click frequency, determine the relative ranks of all HTML-only links, and average them to obtain the mean relative rank of HTML-only links in the respective article.
In the extreme, a mean relative rank of zero (one) implies that the HTML-only links are the most (least) popular out-links of the article.

\begin{figure}[t]
\centering
\includegraphics[width=0.9\columnwidth]{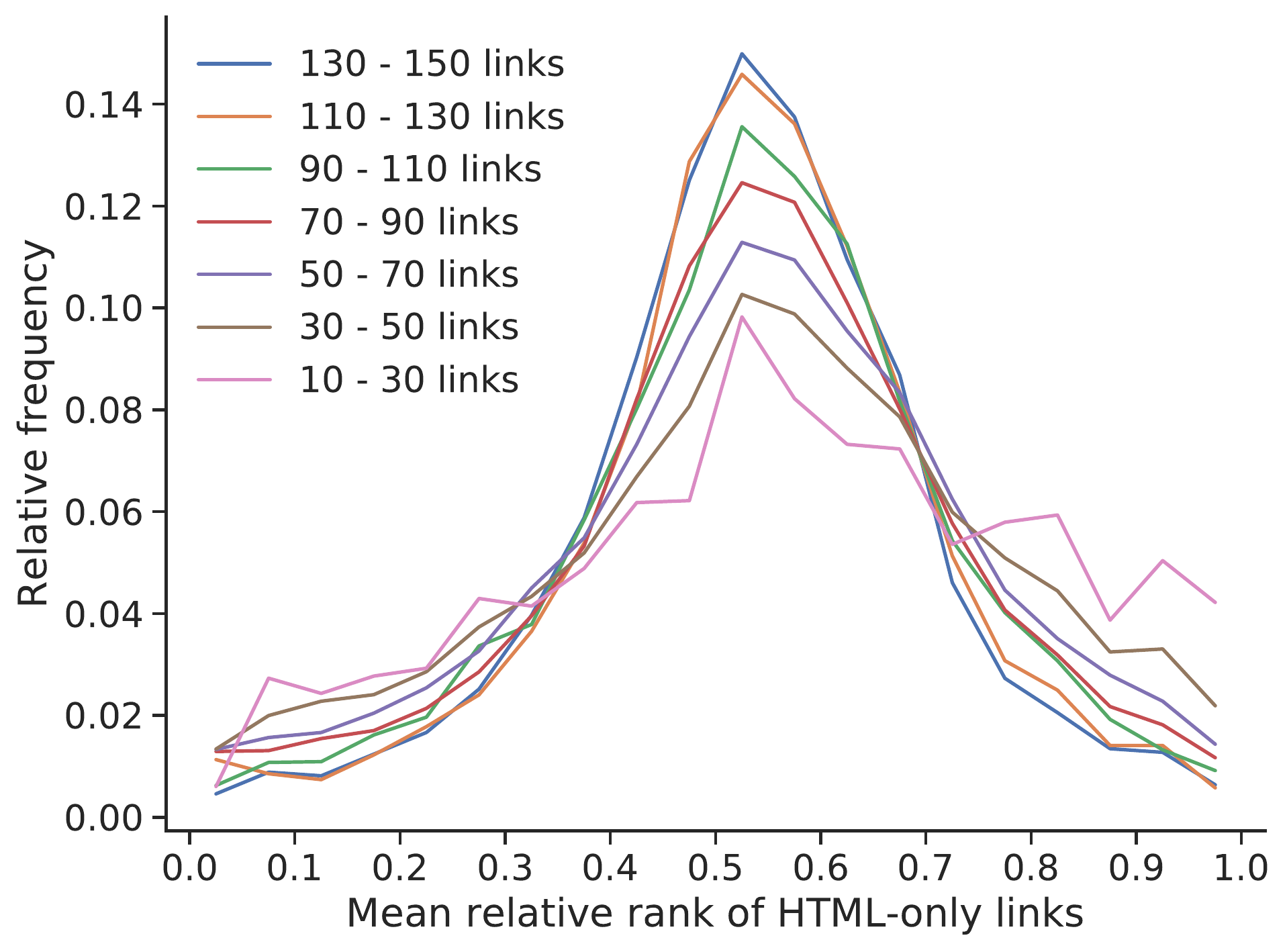} 
\caption{
Histograms of mean relative rank of HTML-only links among all HTML links in terms of click frequency, averaged over 405K articles.
One curve per out\hyp degree bracket.
}
\label{expected_position}
\end{figure}

\Figref{expected_position} shows histograms of the mean relative rank of HTML-only links.
To exclude the total number of out-links as a confound, we stratify articles by the number of out-links and draw a separate histogram per stratum.
If HTML-only links were the least important links, the histograms would show a sharp peak at 1;
if HTML-only links were no different from the other links, the histogram would show a sharp peak at 0.5.
We clearly see that reality resembles the latter case much more than the former case.
From a navigational perspective, HTML-only links are as important as the links also present in wikitext, and to disregard them is to neglect a significant portion of users' interests.

\xhdr{Beyond hyperlinks}
This section illustrated the added value of \name over raw wikitext dumps using the example of hyperlinks, but hyperlinks are not the only information to remain hidden to researchers working with wikitext only.
Templates and modules invoked during the parsing process may also add tables, images, references, and more.

\section{Conclusion}
\label{sec:Conclusion}

To date, \WP's revision history was available only in raw wikitext format, not as the HTML that is produced from the wikitext when a page is requested by clients from the \WP servers.
Since, due to the expansion of templates and modules, the HTML seen by clients tends to contain more information than the raw wikitext sources, researchers working with the official wikitext dumps are studying a mere approximation of the true appearance of articles.

\name solves this problem.
We parsed English \WP's entire revision history from wikitext (nearly 1~TB bzip2\hyp compressed) to HTML (7~TB gzip\hyp compressed) and make the resulting dataset available to the public.

In addition to the data, we also release the code of our custom architecture for parallelized wikitext\hyp to\hyp HTML parsing, hoping that other researchers will find it useful, \eg, for producing HTML versions of \WP's revision history in languages other than English.

% researchers working with the readily available wikitext dumps have had to choose between
% (1)~either working with a mere approximation of what content looked like to readers of \WP or
% (2)~performing wikitext\hyp to\hyp HTML parsing themselves using the \WP API, a process too slow to be performed on the full revision history and which itself produces inaccurate HTML versions of historical revisions.

\balance

\bibliographystyle{aaai} \bibliography{references}

\end{document}